\documentclass[journal,onecolumn]{IEEEtran}
\ifCLASSINFOpdf
  \usepackage[pdftex]{graphicx}
  \graphicspath{ {pictures/}{../pdf/}{../jpeg/} }
  \DeclareGraphicsExtensions{.pdf,.jpeg,.png,.PNG, .tif, .TIF, .eps}
\else
\fi
\makeatletter
\newcommand*{\rom}[1]{\expandafter\@slowromancap\romannumeral #1@}
\makeatother
\usepackage[T1]{fontenc}
\DeclareFontFamily{T1}{calligra}{}
\DeclareFontShape{T1}{calligra}{m}{n}{<->s*[1.44]callig15}{}
\DeclareMathAlphabet\mathcalligra   {T1}{calligra} {m} {n}
\DeclareMathAlphabet\mathzapf       {T1}{pzc} {mb} {it}
\DeclareMathAlphabet\mathchorus     {T1}{qzc} {m} {n}
\DeclareMathAlphabet\mathrsfso{U}{rsfso}{m}{n}
\usepackage[cal=esstix,frak=euler,scr=boondox,bb= pazo]{mathalfa}
\usepackage{yfonts}
\usepackage{cite}
\usepackage{ulem} % For strike through
\usepackage{stfloats}
\usepackage{mathtools}%to have a one-clolumn-width equation in a two coloumn paper
\usepackage{cuted}%to have a one-clolumn-width equation in a two coloumn paper
\usepackage{hyperref} % DO not forget to include the following code to have a blue hyperref
\hypersetup{
    colorlinks=true,
    linkcolor=blue,
    filecolor=magenta,      
    urlcolor=cyan,
    pdftitle={Sharelatex Example},
    bookmarks=true,
    }
\usepackage{xcolor} % to write in color
\usepackage{balance} %to balance the last page columns
\usepackage[utf8]{inputenc}
\usepackage{lmodern}
\usepackage{textcomp}
\usepackage{amsmath }
\DeclareMathOperator*{\minA}{min} % Jan Hlavacek
\DeclareMathOperator*{\argminB}{argmin}   % Jan Hlavacek

\usepackage{mathtools}
\usepackage{physics}
\usepackage{array}
\usepackage{amssymb}
\usepackage{amsthm}
\usepackage{epstopdf} %converting to PDF
\usepackage[ruled,vlined]{algorithm2e}
\usepackage{comment}
\usepackage{amsmath}       
\usepackage{epsf}        
  {
      \theoremstyle{plain}
      
  }
\theoremstyle{plain}

\theoremstyle{definition}

\theoremstyle{remark}

\usepackage{flushend}
\usepackage{tabularx,booktabs,textcomp}
\newcolumntype{C}{>{\centering\arraybackslash}X} % centered version of "X" type
\newcolumntype{L}{>{\raggedright\arraybackslash}X} % LEFT version... "\...right" works?

\begin{document}
%
% paper title
% Titles are generally capitalized except for words such as a, an, and, as,
% at, but, by, for, in, nor, of, on, or, the, to and up, which are usually
% not capitalized unless they are the first or last word of the title.
% Linebreaks \\ can be used within to get better formatting as desired.
% Do not put math or special symbols in the title.
\title {Optimization of Power Control for Autonomous Hybrid Electric Vehicles with Flexible Power Demand}
%
%
% author names and IEEE memberships
% note positions of commas and nonbreaking spaces ( ~ ) LaTeX will not break
% a structure at a ~ so this keeps an author's name from being broken across
% two lines.
% use \thanks{} to gain access to the first footnote area
% a separate \thanks must be used for each paragraph as LaTeX2e's \thanks
% was not built to handle multiple paragraphs
%
%
\author{Mohammadali~Kargar,
	     Xingyong~Song
        % <-this % stops a space
\thanks{A preliminary work was presented in a conference \cite{kargar2022power}.  The journal paper is significantly different from the conference version.  More detailed algorithm derivation and explanation, adaptation of rechable set for ADP training, and more testing results are presented in the submitted journal paper. In addition, major rewordings are conducted to avoid being repetitive.}
\thanks{M. Kargar is with the Department of Mechanical Engineering, College of Engineering, Texas A\&M University, College Station, TX 77843, USA. (e-mail:
mohammadkrg@tamu.edu).}% <-this % stops a space
\thanks{X. Song is with the Department of Engineering Technology and Industrial
Distribution, the Department of Mechanical Engineering, and the Department of Electrical and Computer Engineering, College of Engineering, Texas A\&M University, College Station, TX 77843, USA. (e-mail:
songxy@tamu.edu).}% <-this % stops a space
%\thanks{Manuscript received mm dd, 20XX.}
}

\maketitle

% As a general rule, do not put math, special symbols or citations
% in the abstract or keywords.
\begin{abstract}
Technology advancement for on-road vehicles has gained significant momentum in the past decades, particularly in the field of vehicle automation and powertrain electrification. The optimization of powertrain controls for autonomous vehicles typically involves a separated consideration of the vehicle's external dynamics and powertrain dynamics, with one key aspect often overlooked. This aspect, known as \textit{flexible power demand}, recognizes that the powertrain control system does not necessarily have to precisely match the power requested by the vehicle motion controller at all times. Leveraging this feature can lead to control designs achieving improved fuel economy by adding an extra degree of freedom to the powertrain control. The present research investigates the use of an Approximate Dynamic Programming (ADP) approach to develop a powertrain controller, which takes into account the flexibility in power demand within the ADP framework. The formulation is based on an autonomous hybrid electric vehicle (HEV), while the methodology can also be applied to other types of vehicles. It is also found that necessary customization of the ADP algorithm is needed for this particular control problem to prevent convergence issues. Finally, a case study is presented to evaluate the effectiveness of the investigated method.
\end{abstract}

% Note that keywords are not normally used for peerreview papers.

% For peer review papers, you can put extra information on the cover
% page as needed:
% \ifCLASSOPTIONpeerreview
% \begin{center} \bfseries EDICS Category: 3-BBND \end{center}
% \fi
%
% For peerreview papers, this IEEEtran command inserts a page break and
% creates the second title. It will be ignored for other modes.
\IEEEpeerreviewmaketitle

\section{Introduction}  
% The very first letter is a 2 line initial drop letter followed
% by the rest of the first word in caps.
% 
% form to use if the first word consists of a single letter:
% \IEEEPARstart{A}{demo} file is ....
% 
% form to use if you need the single drop letter followed by
% normal text (unknown if ever used by the IEEE):
% \IEEEPARstart{A}{}demo file is ....
% 
% Some journals put the first two words in caps:
% \IEEEPARstart{T}{his demo} file is ....
% 
% Here we have the typical use of a "T" for an initial drop letter
% and "HIS" in caps to complete the first word.
\IEEEPARstart{T}{ransportation} accounts for over 70 percent of the total oil consumption in the United States, and almost 65 percent of the U.S transportation consumption is from passenger vehicles \cite{qu2020jointly}. The need for lower fuel consumption and cleaner powertrain operation has been driving continuous development in automotive technologies, specifically in the field of powertrain electrification/hybridization \cite{zhang2020energy}. Meanwhile, research on autonomous vehicles has gained major momentum recently, and a much higher percentage of autonomous/semi-autonomous vehicles is expected to be on the road in the near future \cite{kavas2020effects, duarte2018impact}. Although each technology has its own advantages, combining powertrain electrification/hybridization and autonomy has the potential to significantly enhance the fuel efficiency of vehicles.

\begin{figure}[t] % t,h, and b mean "top," "in-line," and "bottom" respectively
  \centering
  \includegraphics[draft=false, width=3.0in]{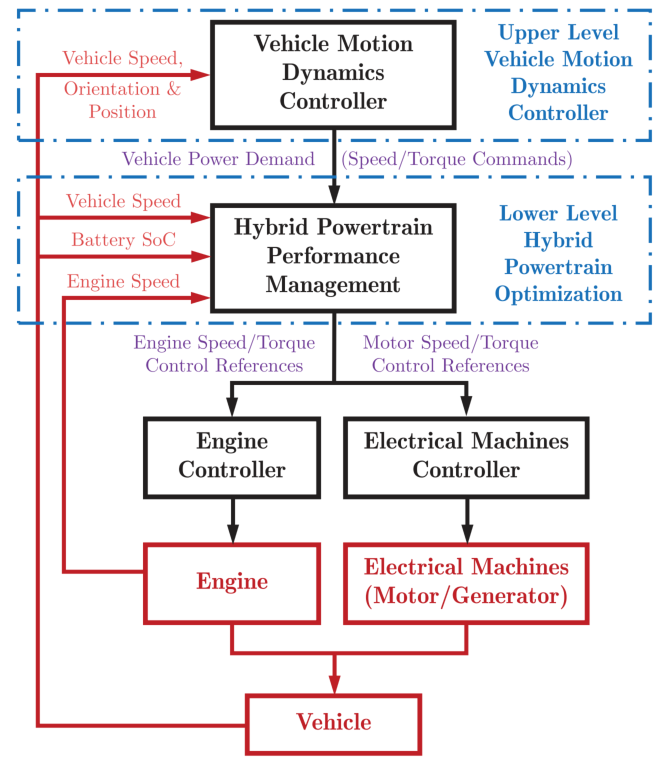}\\
  \caption{Two Levels of Control for Autonomous HEVs.\cite{ghasemi2018powertrain}}\label{fig:Levels}
\end{figure}

Hybridization is an intermediate step on the path toward full electrification \cite{doshi2017hybridization}. An HEV is propelled by an Internal Combustion Engine (ICE) and a battery pack that interacts with electric motors \cite{liu2008modeling}. The presence of an alternative power source gives the powertrain an extra degree of freedom, and thus, the engine can be controlled to operate close to its optimal region \cite{hong2015motor}. The deficit or surplus of the engine power is taken from or stored in the battery pack. The task of optimally supplying the requested power from the power sources is called powertrain energy management or power-split management \cite{borhan2009predictive}. To solve this optimization problem, several methodologies have been applied in the literature, including Dynamic Programming (DP) \cite{wang2012dynamic, perez2006optimization}, model predictive control \cite{zeng2015parallel, huang2017model}, rule-based methods \cite{hofman2007rule, jalil1997rule}, equivalent consumption minimization strategy \cite{paganelli2002equivalent, vskugor2014design}, Pontryagin’s minimum principle \cite{yuan2013comparative, jeong2014development, ahmadizadeh2017energy}, and reinforcement learning \cite{lian2020rule, wu2018continuous, yazar2023actor}. Powertrain energy management is a crucial task since weak power management can result in battery overcharge, battery drainage, and eventually poor fuel economy.

Moreover, Artificial Intelligence (AI) whose versatility is evident in its deployment in many areas including health \cite{boroujeni2021novel, ochieze2023wearable, nasrabadi2022new}, bio treatment \cite{jebellat2023reinforcement, jebellat2021training, ghadermazi2023microbial}, facial recognition \cite{rajoli2023triplet, mehrabi2021age}, environmental conservation \cite{boroujeni2024ic}, safety \cite{pourghorban2023target, mokari2023resilient, mokari2021attack}, aerial transportation \cite{lori2023optimal, ashkoofaraz2022aerial}, control synthesis \cite{samanipour2023stability}, and predicting mechanical properties \cite{safavigerdini2023predicting}, has significantly enhanced the accessibility and capabilities of autonomous vehicles and has made their full potential more achievable than ever before.

Proper control of autonomous vehicles has the potential to dramatically reduce congestion, injuries and fuel consumption \cite{fagnant2015preparing, rashid2023evaluation, martinez2018autonomous}. For example, AVs can have shorter gaps with their leading vehicles which results in improving traffic throughput \cite{litman2020autonomous, ye2019evaluating}. Also, researchers have shown AVs can improve fuel consumption by eliminating the stop-and-go waves in traffics \cite{stern2018dissipation}. Combining autonomy and powertrain hybridization results in an autonomous HEV. As shown in Fig. \ref{fig:Levels}, an autonomous HEV has two levels of control: an upper-level controller and a lower-level controller. The upper-level controller is in charge of optimizing the external dynamics of the vehicle and decides how much driveline torque is needed to meet the maneuvering goals. The lower-level controller is responsible for efficiently allocating the requested driving torque between the ICE and the electric power source.

Most existing research studies these two levels separately: some research studies focus only on the lower-level control design (powertrain energy management) \cite{panday2014review}, \cite{zulkefli2014hybrid}, \cite{kim2011optimal, kim2016comprehensive, kim2015feasibility,  kim2011jump} and some solely studied the upper-level control design (motion tracking/coordination) \cite{anderson2010optimal, foderaro2014distributed, jantapremjit2007control, zulkefli2014hybrid, ma2015lqr, zhang2014distributed, zhang2014leader, yao2020control}. Studies from the U.S. Department of Energy have shown that augmenting these two optimization problems can offer fuel-saving potentials that cannot be achieved by powertrain optimization alone \cite{atkinson2015powertrain, kargar2023integratedACC}. The underlying reason is that when these levels are solved separately, the upper-level controller cannot observe the dynamics of the powertrain, and thus, the requested driving power may not consider the powertrain's most efficient condition.
%There are some research works that have studied the fuel economy of autonomous HEVs considering both vehicle coordination and powertrain optimization together. However, because co-optimizing these two problems are computationally hard, they usually have considered the hierarchical control strategy [ref]. In this strategy, they first solve the coordination problem of the vehicle and then solve for the hybrid powertrain optimization [ref: Masood – ref: Langari]. This strategy suffers from the fact that without embedding the complex hybrid powertrain dynamics into the coordination control design, the significant fuel-saving potential offered by the hybrid propulsion architecture cannot be fully achieved. 

The benefits of joint optimization of these two levels have urged the researchers to focus on solving the speed control and power energy management in a unified framework \cite{zhang2020real, wang2022multi,zhao2019optimal, mahbub2019concurrent, ma2017integrated}. To mitigate the computational burden due to the high coupling between the dynamics of the upper level and lower level, \cite{zhang2020real, wang2022multi,zhao2019optimal, mahbub2019concurrent} proposed a hierarchical control framework to optimize (1) the vehicle’s speed profile, and (2) the powertrain efficiency of the vehicle for the optimal speed profile derived in (1). Although these studies have shown promising results, these two levels are still solved separately, and the full potential of the integrated optimization in fuel minimization is still unleashed. In \cite{ma2017integrated}, MA \textit{et al.} devised a control architecture to solve the vehicle dynamics and the powertrain dynamics in one optimization problem. Nevertheless, the drawback in their methodology (forward DP) is that it depends on the system's initial condition and new optimization is needed anytime the initial condition changes \cite{engbroks2018applying}. 

This paper will explore a customized approach in which still two levels are optimized separately and the lower-level controller receives the power demand from the upper-level controller. However, the lower-level controller is given an extra degree of freedom by using the unique property in autonomous HEVs, i.e., flexibility in power demand. This property implies that the vehicle does not have to precisely match the power demanded by the upper-level controller in real-time. There can be some variations in power that deviate from what is required by the upper-level controller. This feature is only achievable in autonomous HEVs where both upper-level and lower-level controllers operate in the background, and no human driver is intervening. This approach has been sought by some researchers recently \cite{ghasemi2018powertrain, zhang2021adaptive} which has shown promising results in terms of fuel consumption reduction. However, the methodologies they have used make them less practical in the application of HEV power management. In \cite{ghasemi2018powertrain}, Pontryagin’s minimum principle (PMP) was used to optimize the powertrain control optimization. For conventional HEVs, PMP is a powerful method for addressing powertrain energy management. However, when constraints on states exist in the problem, it can be challenging to implement PMP to meet them \cite{sanchez2020energy, serrao2009ecms}. Thus, for autonomous HEVs with flexible driveline power demand wherein state constraints are applied \cite{ghasemi2018powertrain}, PMP can cause some convergence issues.
Also, \cite{zhang2021adaptive} deployed an adaptive equivalent consumption minimization strategy to optimize power splits for an automated parallel HEV with flexible power demand. However, the main drawback in \cite{zhang2021adaptive} is that the proposed method could not satisfy the final conditions.

The research presented in this paper investigates a customized ADP approach, which is used for the first time to address the issue of powertrain optimization with flexible power demand. Solving this problem using ADP is independent of the initial condition of the system and requires much less memory storage capacity compared to the conventional DP. However, directly using standard ADP to solve the optimal control problem can be challenging. The standard ADP method needs the dynamical system to have a quadratic cost function to minimize, as well as an affine structure of the inputs. However, these conditions do not hold in the energy management of HEVs. Also, the system has nonlinear constraints on states and control inputs which standard ADP cannot take care of. Therefore, two methods are adopted to address these challenges. Initially, the ADP method is modified to facilitate its use in handling non-quadratic cost functions with non-affine inputs. Second, the concept of reachable set \cite{elbert2012implementation} is adopted in implementing ADP to meet the nonlinear constraints on states and control inputs. Studies have shown that utilizing the concept of reachable sets can improve the  accuracy and the efficiency of the optimization solution noticeably \cite{elbert2012implementation}.

The outline of this paper is as follows. In Section \ref{Dynamics}, the HEV modeling is discussed. Reachable sets and ADP framework are introduced in Section \ref{Optimal_Control}, followed by a numerical example in Section \ref{Results}. At last, concluding remarks are given in Section \ref{Conclusion}.  

\section{HEV Dynamics} \label{Dynamics}
\subsection{Upper-Level Dynamics}
\begin{figure}[t]

 \center

  \includegraphics[draft=false, width=3.8in]{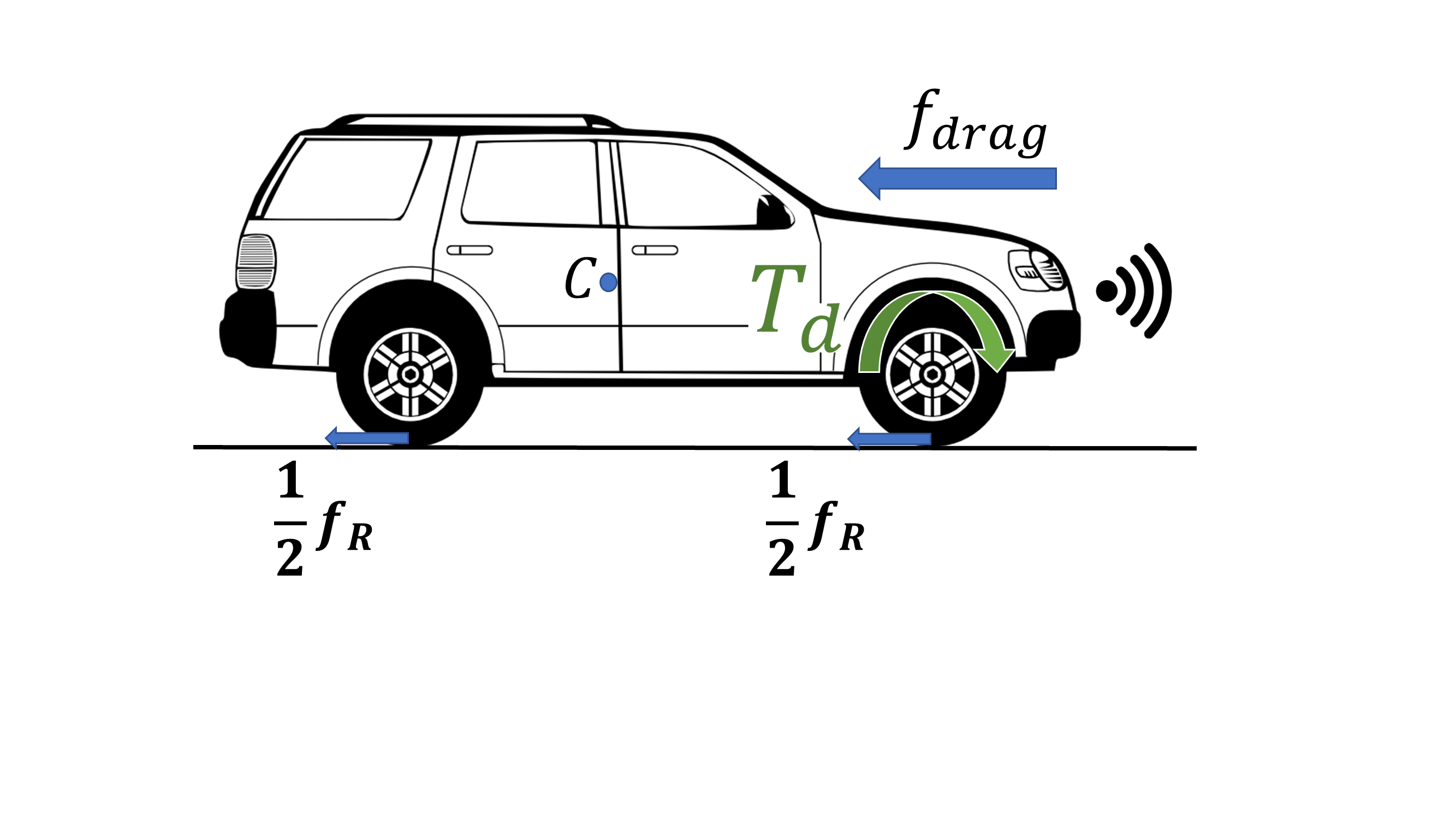}\\
  \caption{Free Body Diagram of the Vehicle.}\label{fig:FBD}
\end{figure}
Fig. \ref{fig:FBD} illustrates the free-body diagram of a vehicle moving in a straight path within an inertial frame of reference. Note that the reaction forces on each individual wheel are summed up at their mid-axles (bicycle model). Let $x$ represent the longitudinal position of the center of mass of the vehicle $C$, and $v$ denote its longitudinal velocity, the external kinematics and dynamics of the vehicle can be expressed as follows:
\begin{align}
    \label{eq:xdot}
	\dot{x}(t)		&=v(t) ,  &x(0)= x_{0}\\
	\dot{{v}}(t)   &= \frac{1}{{m}} \left[\frac{1}{r}{{T}_{d}(t)}-{f_{drag}} -{f_{R}}\right],   &v(0)= v_{0}
	\nonumber 
	\\
	{f_{drag}}  &= \frac{1}{2}\rho{C_{drag}}{A_{f}}{v}(t)^{2}
	\nonumber
	\\	
	{f_{R}}  &= {{\mu}_{R}}{m}{g}.
	\label{eq:vdot}
\end{align}
Here, $m$, $r$, $A_f$, ${f_{drag}}$, ${f_{R}}$, and ${\mu}_{R}$ represent the vehicle's mass, wheel radius, effective frontal area, drag force, rolling resistance force, and rolling resistance coefficient, respectively. Additionally, $\rho$ and $C_{drag}$ correspond to the air density and coefficient of drag, respectively.\\
\subsection{Lower-Level Dynamics}
\subsubsection{Powertrain Dynamics}

\begin{figure}[t] % t,h, and b mean "top," "in-line," and "bottom" respectively
  \centering
  \includegraphics[draft=false, width=3.4in]{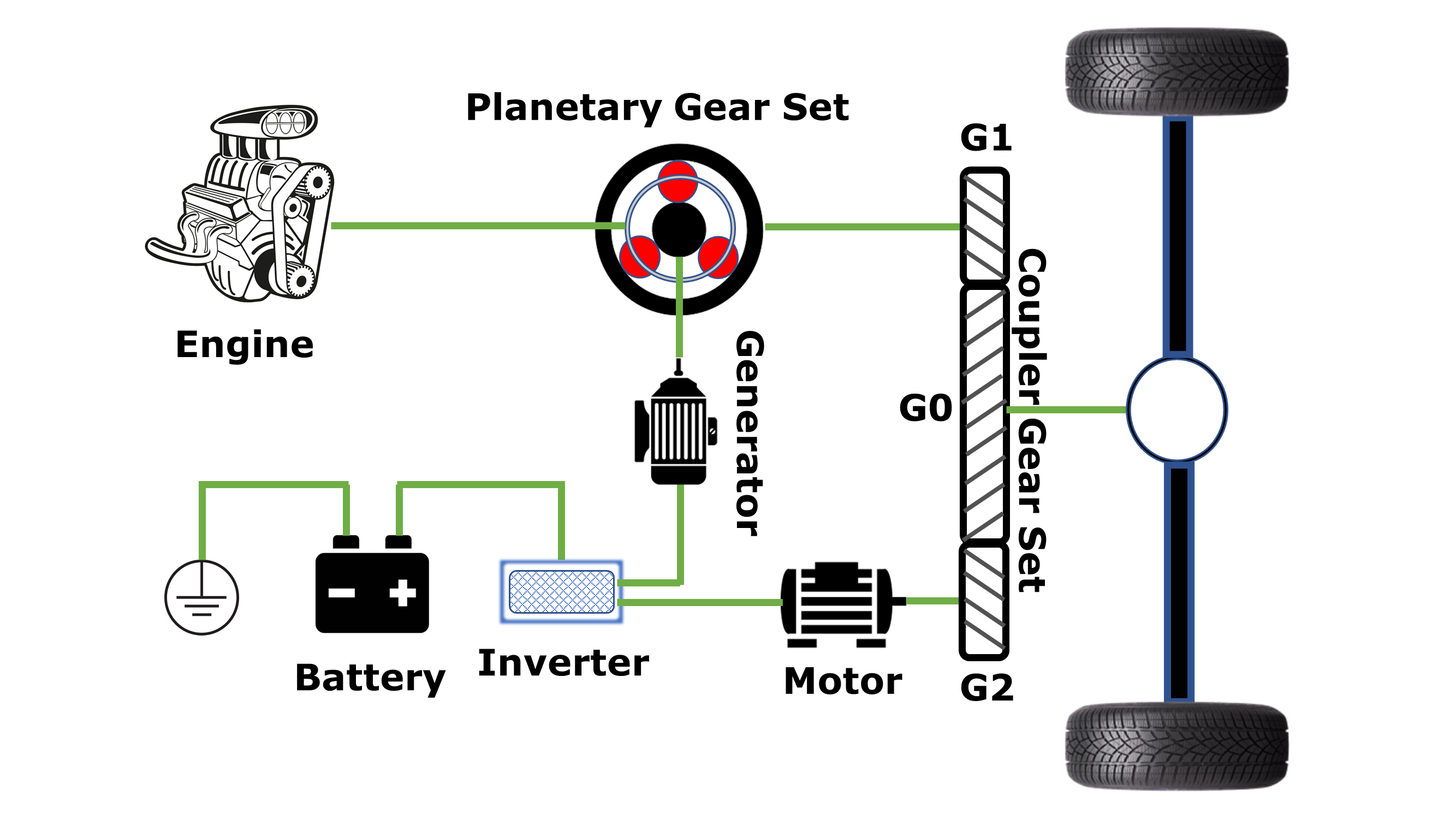}\\
  \caption{THS Powertrain Schematic.}\label{fig:Powertrain}
\end{figure}

In this study, a power-split hybrid powertrain is considered (Fig. \ref{fig:Powertrain}). This mechanism which was commercialized by Toyota is known as Toyota Hybrid System (THS). THS is formed of a battery pack, an ICE, a coupler gear set, a planetary gear set, an inverter, and two electric machines. In the literature, the electric machine that mostly acts as motion power output is labeled as the "motor," and the one that is expected to mostly operate as an electricity generation unit is called the "generator." The powertrain splits the power between the ICE and the battery through the planetary gear set that consists of three main elements: the sun, the carrier, and the ring. The engine is linked to the carrier, and the sun is connected to the generator. 

Gear G0 merges the ring’s power and the motor’s power through two identical gears, G1 and G2, to drive the driveline. Neglecting the inertia of the moving parts in the powertrain and using the power balance at the planetary and the coupler gear sets, the following algebraic equations between different components hold:
\begin{align}
	\omega_{e}(t)		
	& = 
	(
	\frac
	{r_r}
	{r_s+r_r}
	) 
	\omega_{r}(t) 
	+ 
	(
	\frac
	{r_s}
	{r_s+r_r}
	)
	 \omega_{g}(t) \label{eq:we}
	\\
	\omega_{r}(t)		
	& = 
	\omega_{m}(t) \label{eq:wr}
	\\
	\omega_{m}(t)		
	& = 
	k_C \omega_{d}(t) \label{eq:wm}
	\\
	T_{g}(t)		
	&=  
	- (
	\frac
	{r_s}
	{r_s+r_r}
	)
	T_e(t) \label{eq:Tg}
	\\
	T_{r}(t)		
	&=  
	(
	\frac
	{r_r}
	{r_s+r_r}
	)
	T_e(t) \label{eq:Tr}
	\\	
	\frac
	{1}
	{k_C}
	T_{d}(t)
	&=
	T_{m}(t)	
	+	 
	T_{r}(t)					\label{eq:Td}
\end{align}
where $\omega_{e}$, $\omega_{r}$, $\omega_{g}$, $\omega_{m}$, and $\omega_{d}$ denote the angular velocity of the engine, the ring, the generator, the motor, and the driveline, respectively. Likewise, $T_{g}$, $T_{e}$, $T_{r}$, $T_{d}$, and $T_{m}$ represent the torque of the generator, the engine, the ring, the driveline, and the motor, respectively. Also, $r_{r}$, $r_{s}$, and $k_C$ denote the radius of the ring gear, the radius of the sun gear, and the final gear ratio at the coupler gear set, respectively.

Lastly, considering the power balance at the inverter reveals  the following algebraic equation:
\begin{equation}
	P_{batt}(t) 
	= 
	P_{m}(t) 
	+
	P_{g}(t)
	=
	 \mu_{m}
	^
	{k_m}
	T_{m}(t)
	\omega_{m}(t) 
	+ 
	\mu_{g}
	^
	{k_g}
	T_{g}(t)
	\omega_{g}(t) 
	\label{eq:Pbatt1}
\end{equation}
where $P_{batt}$, $P_m$, and $P_g$ denote the battery power,  the motor power, and the generator power, respectively. Note that the motor power and generator power can be positive (when they are operating as a motor) or negative (when they are operating as a generator). Parameters $\mu_m$ and  $\mu_g$, respectively, denote the coefficients of efficiency for the motor and the generator when they operate as electricity-producing units. These coefficients vary between 0 and 1. Therefore, $k_m$ and $k_g$ are either equal to -1 when their corresponding electrical machines produce motion power or they are equal to 1 otherwise. Also, a positive $P_{batt}$ means the battery is discharging while a negative $P_{batt}$ shows it is charging.
Using Eqs. (\ref{eq:we}) to (\ref{eq:Pbatt1}), $P_{batt}$ is formulated as:
\begin{equation}
\begin{split}
P_{batt}(t) 
	=
	 &
	\mu_{m}
	^
	{k_m}
	T_{d}(t)
	\omega_{d}(t)
	 - 
	\mu_{g}
	^
	{k_g} 
	T_{e}(t)
	\omega_{e}(t)
	\\
	&
	-
	\frac
	{
	k_{c}
	r_{r}
	(
	\mu_m
	^
	{k_m}
	-
	\mu_{g}
	^
	{k_g}
	)
	}
	{r (r_s+r_r)}
	T_{e}(t)
	{v}(t).
	 \label{eq:Pbatt2}
\end{split}
\end{equation}
\subsubsection{Battery Dynamics}
The dynamics of the battery are modeled using an equivalent circuit model \cite{kargar2023integrated}:
\begin{equation}
\begin{split}
\dot{SoC}(t)= -\frac{V_{batt}-\sqrt{V_{batt}^2-4R_{batt}P_{batt}(t)}}{2R_{batt}Q_{batt}},&\\ SoC(0)= SoC_0& \label{eq:SoC}
\end{split}
\end{equation}
here, state of the charge, open-circuit voltage, internal resistance, and capacitance of the battery are represented by $SoC$, $V_{batt}$, $R_{batt}$, and $Q_{batt}$, respectively. Note that the nominal operating range of $SoC$ is around 40\%–80\% in charge-sustaining operations, where the initial and final values of $SoC$ are equal. For this operating range, the battery’s parameters are almost constant \cite{kim2010optimal}.

\subsubsection{Fuel Consumption Dynamics}
\begin{figure}[!h] % t,h, and b mean "top," "in-line," and "bottom" respectively
  \centering
  \includegraphics[draft=false, width=3.8in]{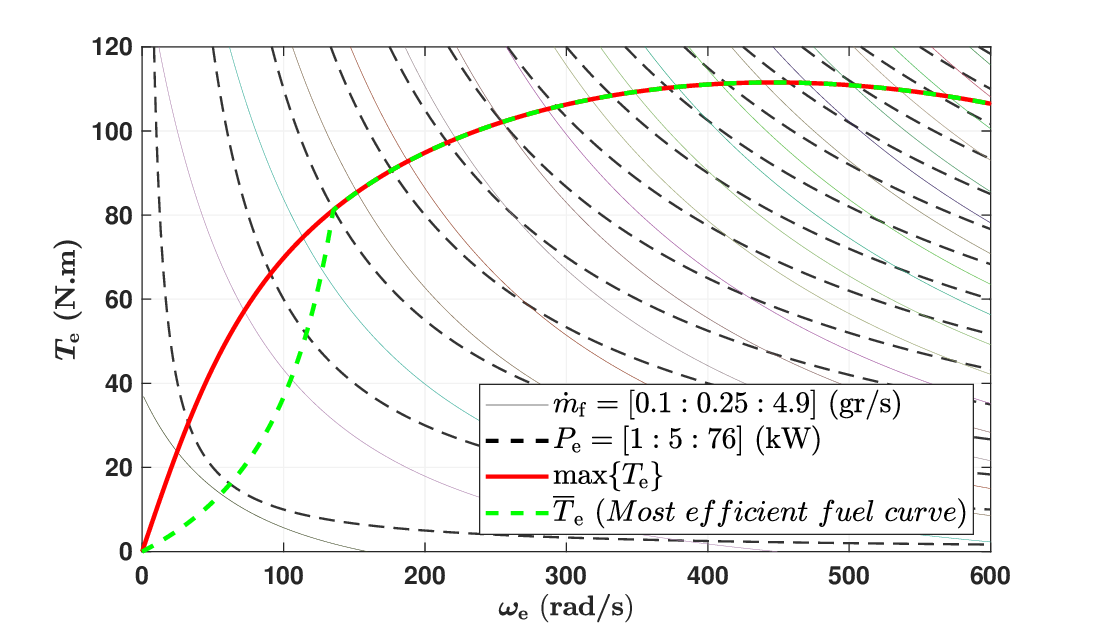}\\
  \caption{Engine Map.}\label{fig:MAP}
\end{figure}
The fuel consumption dynamics are generally governed by the engine's angular velocity ${\omega}_{e}$ and the engine's torque ${T}_{e}$ as depicted by Eq. (\ref{eq:mdotfuel}). 
\begin{equation} \label{eq:mdotfuel}
	\dot
	{m}_{fuel}
	=
	\gamma
	(
	{\omega}_{e}
	,	
	{T}_{e}
	)
\end{equation}
where $\gamma :\mathbb{R}_{+} \times \mathbb{R}_{+}  \rightarrow \mathbb{R}_{+}$ is a mapping whose inputs are ${\omega}_{e}$ and ${T}_{e}$ and measures the fuel consumption rate. $\mathbb{R}_{+}$ depicts the set of nonnegative real numbers. An engine map generated from the experimental data is typically employed to capture this correlation, as illustrated in Fig. \ref{fig:MAP}. The solid red line and dash green line depict the maximum engine torque and the optimal engine torque $\bar{T}_{e}$ for different values of ${\omega}_{e}$, respectively. Given the two degrees of freedom in the powertrain for power supply, it is a reasonable assumption \cite{zhang2021adaptive, ma2017integrated} that for any engine power $P_e(t) = T_e(t) * \omega_e(t)$, the corresponding solution pair $(T_e(t), \omega_e(t))$ will position the engine power at its optimal efficiency point on the engine map. $\bar{T}_{e}$ for this point is mathematically approximated by the following equation:
\begin{align} \label{eq:Te_opt}
	\bar{T}_{e}
	&=
	(60 * atan (\frac{{\omega}_{e}}{70})) - 0.00018 * {{\omega}_{e}}^{2}+0.14*{\omega}_{e}
\end{align}

\subsubsection{Flexible Power Demand Dynamics}
\begin{figure}[!t]

 \center

  \includegraphics[draft=false, width=3.8in]{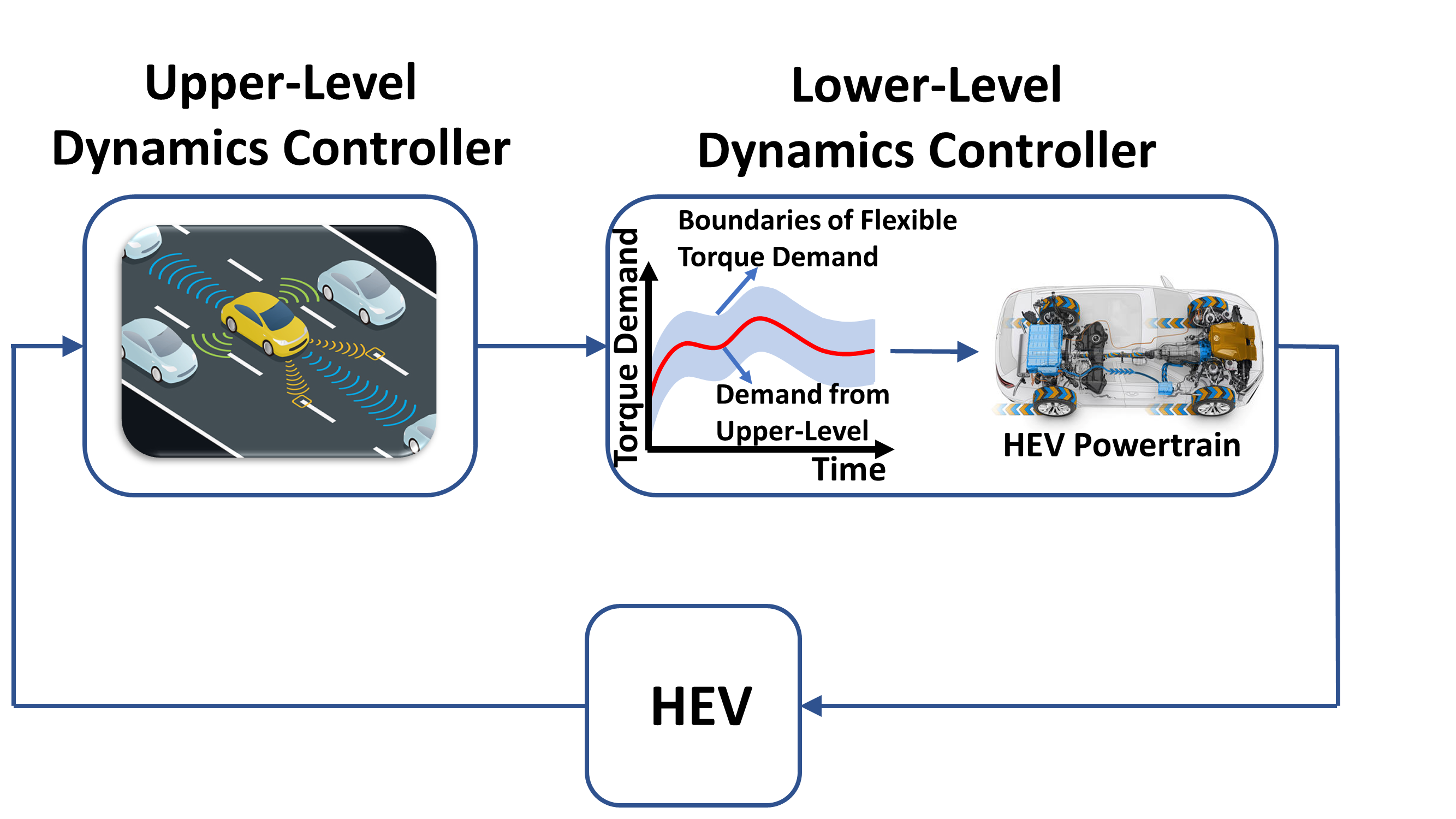}\\
  \caption{Energy Management Hierarchy with Flexible Torque Demand.}\label{fig:FlexibleTorque}
\end{figure}

As stated before, the upper-level controller determines the torque and the power required by the vehicle based on the upper-level dynamics. Conventionally, these power and torque demands are fed to the lower-level controller. The lower-level controller determines the distribution of power among the ICE, electrical machines, and the battery pack to meet the power demand.
In the majority of the research studies, the driveline power demand is strictly met by the lower-level controller. However, in an autonomous HEV, there exists a uniqueness which allows the lower-level controller to have certain degrees of flexibility to meet the instantaneous power requested by the upper level. This flexibility in power demand which is illustrated in Fig. \ref{fig:FlexibleTorque}, adds an extra degree of freedom to the powertrain control optimization which can further improve fuel efficiency.

Variations in the power supplied by the powertrain will lead to deviations in the anticipated acceleration determined by the upper-level controller. Consequently, there will be corresponding deviations in the expected velocity and displacement of the vehicle. However, in order to ensure that the vehicle reaches its intended destination, allowances for velocity and displacement deviations are made only during intermediate steps. In other words, although the longitudinal displacement, velocity, and driveline torque demand may differ from their expected values at each intermediate moment determined by the upper-level controller, the deviation in the longitudinal displacement and velocity must diminish as time approaches the end of the considered time horizon (note that the time horizon can be a short period, and the optimization can be done for each time period going forward one by one). 

Denoting the flexible driveline torque, longitudinal displacement, and velocity as $\tilde{T}_{d}$, $\tilde{x}$, and $\tilde{v}$ respectively, the external dynamics of the vehicle can be rewritten to incorporate the aforementioned flexibilities as follows:

\begin{align}
    \label{eq:xbardot}
	\dot{\tilde x}(t)	&=\tilde v(t) , \qquad\qquad\qquad \tilde x(0)= x(0)= x_{0}\\
	\dot{{\tilde v}}(t)  & = \frac{1}{{m}} \left[\frac{1}{r}{{\tilde T}_{d}(t)}-{f_{drag}} -{f_{R}}\right],  \tilde v(0)= v(0)= v_{0}
	\nonumber 
	\\
	{f_{drag}} & = \frac{1}{2}\rho{C_{drag}}{A_{f}}{\tilde v}(t)^{2}.
	\label{eq:vbardot}
\end{align}
It is worth noting that $\tilde x(0)= x(0)$ and $\tilde v(0)= v(0)$ are valid assumptions based on the fact that at the start of the driving cycle, the lower-level controller begins with the same initial longitudinal displacement and velocity as determined by the upper level.

Define $\Delta x$ as the difference between the flexible longitudinal displacement and the longitudinal displacement obtained from the upper-level controller as below:
\begin{equation} \label{eq:deltax_def}
\Delta x \triangleq \tilde{x} - x.
\end{equation}
Also, $\Delta v$ is defined in a similar way:
\begin{equation} \label{eq:deltav_def}
\Delta v \triangleq \tilde{v} - v
\end{equation}
By taking into account the equations (\ref{eq:xdot}), (\ref{eq:vdot}), (\ref{eq:xbardot}), and (\ref{eq:vbardot}), the following relationships can be derived:
\begin{equation}\label{eq:deltaxdot}
    \Delta\dot{{x}}(t) =\Delta{v}(t),\qquad\qquad \Delta{x}(0) = 0
\end{equation}
\begin{equation}
\begin{split}
    \label{eq:deltavdot}
    \Delta\dot{{v}}(t) & = \dot{\tilde{v}}(t) - \dot{v}(t)= \frac{1}{{m}} [-\frac{1}{2}\rho{C_{drag}}{A_{f}}\Delta{v}(t) (2{v(t)}\\
	&+\Delta{v}(t)) +\frac{1}{r}{\Delta{T}_{d}(t)}] , \quad\qquad\qquad \Delta{v}(0) = 0.
\end{split}
\end{equation}
where $\Delta{T}_{d}\triangleq \tilde{T_{d}} - T_{d}$ denotes the amount of flexibility in the torque demand and serves as a control input. 
\subsection{State Space Model} \label{State_Space}
In order to monitor the $SoC$ of the battery and the deviations in longitudinal displacement and velocity, we define the following state vector:
\begin{equation} 
\textbf{X}(t) \triangleq {[\Delta{x},\Delta{v}, SoC]}^T. \label{eq:state_vector}
\end{equation}
Similarly, the vector of inputs is defined as follows:
\begin{equation} 
\textbf{U}(t) \triangleq {[\omega_{eng}, Te_{eng}, \Delta{T_{d}}]}^T. \label{eq:input_vector_extended}
\end{equation}
By utilizing equation (\ref{eq:Te_opt}), the input vector can be simplified to:
\begin{equation} 
\textbf{U}(t) = {[\omega_{eng},  \Delta{T_{d}}]}^T. \label{eq:input_vector}
\end{equation}
Taking into account equations (\ref{eq:Pbatt2}), (\ref{eq:SoC}), (\ref{eq:deltaxdot}), and (\ref{eq:deltavdot}), the system dynamics can be summarized as:
\begin{equation}
\dot{\textbf{X}}(t) = \mathscr{F}(\textbf{X}(t), \textbf{U}(t)), \qquad \textbf{X}(0)=\textbf{X}_0 \label{eq:XDOT}
\end{equation}
where $\mathscr{F}$ is specified as below:
\begin{equation}\label{eq:F}
   %\left[
%	\bgin{gathered}  
	\mathscr{F}
	\triangleq
	 %\left [
%	\begin{gathered}
	\begin{bmatrix}
	\Delta{v}
	\\
	\frac
	{1}
	{m}
	\left
	[	
	-\frac
	{1}
	{2}
	\rho
	{C_{drag}}
	{A_{f}}
	\Delta{v}(2v + \Delta{v})
	+
	\frac
	{1}
	{r}
	\Delta{T_{d}}
	\right
	]
	\\
	-\frac
	{
	V_{batt}
	-\sqrt
	{
	V_{batt}^2
	-4
	R_{batt}
	P_{batt}(t)
	}
	}
	{
	2
	R_{batt}
	Q_{batt}
	}
%	\end{gathered}
	%\right]
	\end{bmatrix}
	.
\end{equation}
\section{Optimal Control Formulation} \label{Optimal_Control}

Given a general discrete dynamical system with $n$ states and $m$ inputs, depicted as follows:
\begin{equation}\label{eq:X_general}
{\dot{\textbf{x}}(t)} = \mathbb{F}(\textbf{x}(t), \textbf{u}(t)) 
\end{equation}
where $\textbf{x}(t) \in \mathbb{R}^{n}$, and $\textbf{u}(t) \in \mathbb{R}^{m}$ are the state and the input vector of the system, and $\mathbb{F}$ represents the governing dynamical equation of the system. To tackle the optimization problem, it is necessary to establish a cost function. The cost function $J_c$, in its most comprehensive form, can be defined as below:
\begin{equation}
\begin{split} \label{eq:J_c}
J_c & = \int_0^{t_f} \lambda(\textbf{x}(t), \textbf{x}(t)) ~dt+ \psi(\textbf{x}(t_f))
\end{split}
\end{equation}
where, $t_f$ denotes the final time, and $\lambda(\textbf{x}(t), \textbf{u}(t))$ represents the cost associated with intermediate states and inputs. It is important to note that $\textbf{x}(t_f)$ denotes the final state vector, and $\psi(.):\mathbb{R}^{n} \rightarrow \mathbb{R}{+}$ refers to the penalizing function, which is a design parameter chosen as a nonnegative function. The purpose of this function is to ensure that the system reaches the desired terminal point $\textbf{x}_{des}(t_f)$ by penalizing state vectors that deviate significantly from $\textbf{x}_{des}(t_f)$.

By introducing the discretization sample time as $\delta t$ and the discrete time index as $k$, Eq. (\ref{eq:X_general}) can be discretized using the Euler method:

\begin{equation}
\textbf{x}({k+1}) = \textbf{x}({k}) + \delta t \mathbb{F}(\textbf{x}(k), \textbf{u}(k)), \qquad k = 0, 1, 2, ..., N-1. \label{eq:X_discretized}
\end{equation}
Likewise, equation (\ref{eq:J_c}) can be discretized as:
\begin{equation}\label{eq:J_discretized}
	J
	= 
	\sum_{k=0}^{N-1} \delta t \left (\lambda(\textbf{x}(k), \textbf{u}(k)) \right) + \psi(\textbf{x}(N)). 
\end{equation}
Here, $N$ is defined as $\frac {t_f} {\delta t}$. In accordance with the definition of the cost function in Eq. (\ref{eq:J_discretized}), the cost-to-go $V_{k}(\textbf{x}(k))$ is determined as the cost from the state $\textbf{x}(k)$ at time index $k$ to the end of the time horizon, and can be expressed as:

\begin{equation} 
\begin{split}\label{eq:V_X}
V_{k}(\textbf{x}(k)) 
	= 
	&
	\sum_{\tau=k}^{N-1} \delta t \left (\lambda(\textbf{x}(\tau), \textbf{u}(\tau)\right) 	
	+
	\psi(\textbf{x}(N))
	\\
	=
	&
	 \delta t \left ( \lambda(\textbf{x}(k), \textbf{u}(k)\right) 
	+
	\sum_{\tau=k+1}^{N-1} \delta t \left ( \lambda(\textbf{x}(\tau), \textbf{u}(\tau)\right) 	
	\\
	=
	&
	 \delta t \left ( \lambda(\textbf{x}(k), \textbf{u}(k)\right) 
	+
	V_{k+1}(\textbf{x}(k+1)), 
	\\
	&	
	\qquad\qquad\qquad\qquad\qquad k = 0, 1, 2, ..., N-1. 
\end{split}
\end{equation}
This equation straightforwardly indicates that the cost of transitioning from $\textbf{x}(k)$ at time index $k$ to $\textbf{x}(N)$ is equivalent to the cost of transitioning from $\textbf{x}(k)$ to $\textbf{x}(k+1)$ plus the cost of transitioning from $\textbf{x}(k+1)$ to $\textbf{x}(N)$. It is also important to note that:
\begin{equation}\label{eq:V_XN}
V_{N}(\textbf{x}(N)) = \psi(\textbf{x}(N)).
\end{equation}

Let optimal cost-to-go $V_{k}^{*}(\textbf{X}(k))$ be defined as the minimum cost of transitioning from $\textbf{x}(k)$ to $\textbf{x}(N)$. According to the Bellman's principle of optimality \cite{bellman1966dynamic}, $V_{k}^{*}(\textbf{X}(k))$ can be expressed recursively as:

\begin{align}
	&V_{k}^{*}(\textbf{x}(k)) 
	=	
	\minA\limits_
	{\textbf{u}(k)} 
	{
	\big(\delta t \left (\lambda(\textbf{x}(k), \textbf{u}(k)\right) 
	+
	V_{k+1}^{*}(\textbf{x}(k+1))\big)
	}
	\nonumber
	\\
	&V_{N}^{*}(\textbf{x}(N)) 
	=	
	\psi(\textbf{x}(N))		
	\label{eq:Vstar_XN}
\end{align}
\subsection{Standard ADP Methods} \label{Standard}
Assume that the dynamical system in Eq. (\ref{eq:X_discretized}) is control-affine as shown below:
\begin{equation}\label{eq:X_affine}
{\textbf{x}}({k+1}) = \mathscr{f}(\textbf{x}(k)) + \mathscr{g}(\textbf{x}(k))\textbf{u}(k), \qquad k = 0, 1, 2, ..., N-1 
\end{equation}
Also, assume that a quadratic cost function with the intermediate cost is shown as below:
\begin{equation}\label{eq:intermediate_cost}
\lambda(\textbf{x}(k), \textbf{u}(k))  =\frac{1}{2} \textbf{x}(k)^T \mathscr{Q} ~\textbf{x}(k) + \frac{1}{2} \textbf{u}(k)^T \mathscr{R}~ \textbf{u}(k).
\end{equation}
%
%
%
%t
% 
where $\mathscr{f}$ and $\mathscr{g}$ represent the dynamics of the system, and matrices $\mathscr{Q}$ and $\mathscr{R}$ are positive semi-definite, and positive definite matrices, respectively. It can be shown \cite{lewis2009reinforcement} that the optimal control input $\textbf{u}^{*}(\textbf{x}(k))$ satisfies the Bellman optimality condition $\frac{\partial {V^{*}_{k}(\textbf{x}(k)) }}{\partial {\textbf{u}^{*}(\textbf{x}(k))}} = 0$, and can be calculated as:
\begin{figure}[!t]

 \center

  \includegraphics[draft=false, width=4.2in]{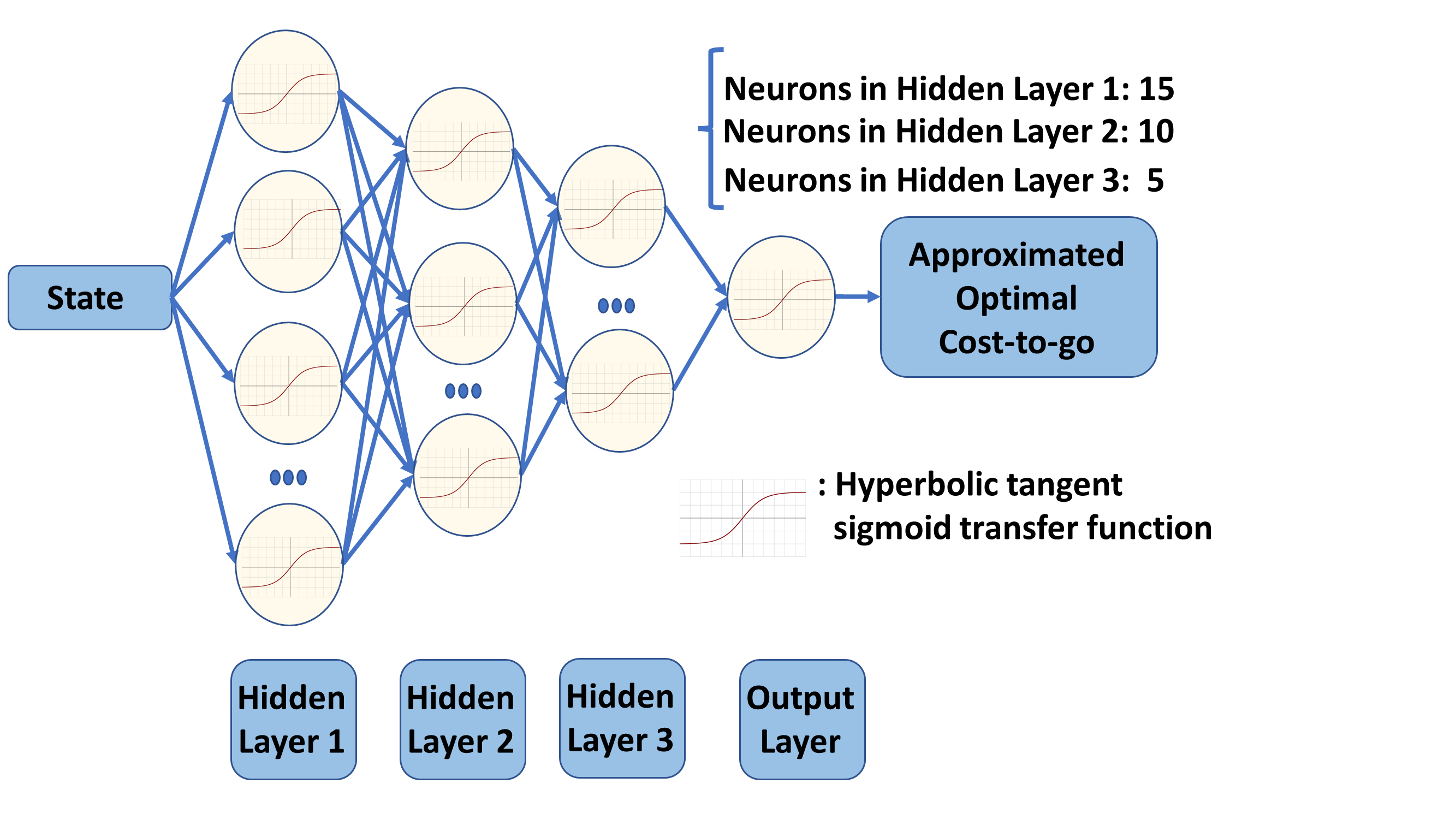}\\
  \caption{Architecture of the Deep Neural Network.}\label{fig:NN}
\end{figure}

\begin{equation} \label{eq:u_star_affine}
\textbf{u}^*(\textbf{x}(k)) = -\mathscr{R}^{-1} \mathscr{g}(\textbf{x}(k))^T \frac{\partial V^{*}_{k+1}(\textbf{x}(k+1))}{\partial \textbf{x}({k+1})}
\end{equation} 
Mainly, there are two types of ADP methods to solve this equation: adaptive critics and Single Neuron Adaptive Critics (SNAC). Although they both use neural networks to find the sub-optimal solution, the main difference between them is that in SNAC, the neural network is used to approximate the derivative of the cost-to-go, i.e., $\frac{\partial V^{*}_{k+1}(\textbf{x}(k+1))}{\partial \textbf{x}({k+1})}$. In adaptive critics, two sets of neural networks are used: one to approximate the optimal cost-to-go $V^{*}_{k}(\textbf{x}(k))$ and one to approximate the optimal action $\textbf{u}^*(\textbf{x}(k))$. Both of these conventional ADP methods are fast and provide suboptimal solutions. However, there are some restrictions with these conventional ADP methods. First, they assume that the system is control input affine \cite{heydari2012finite, li2021novel, kiumarsi2019optimal, abu2005nearly}, and this assumption is the essential assumption needed to solve for the optimal control inputs.  Although there are alternative methods to alleviate this requirement, they typically can induce other challenges. Secondly, the cost function needs to be a quadratic function of the states and the inputs \cite{heydari2012finite, li2021novel, kiumarsi2019optimal}. Another issue in the conventional ADP methods is that they usually solve problems with unconstrained inputs\cite{heydari2012finite, li2021novel, kiumarsi2019optimal}. Note that in Eq. (\ref{eq:u_star_affine}), the inputs found are proportionally dependent on the magnitude of $\frac{\partial V^{*}_{k+1}}{\partial \textbf{x}({k+1})}$, and they can be unreasonably high if not well constrained. 

\subsection{Customized ADP Method} \label{Customized}
With the restrictions explained in \ref{Standard}, standard ADP methods cannot be directly applied to the HEV power control optimization problem considered in this paper. Firstly, the system described in Eq. (\ref{eq:XDOT}) is highly non-affine due to the battery dynamics. Secondly, in order to minimize fuel consumption throughout the drive cycle, the intermediate cost is defined as follows:
\begin{equation}\label{eq:intermediate_cost}
\lambda(\textbf{X}(t), \textbf{U}(t))  =\dot{m}_{fuel}(t).
\end{equation}
Thus, the cost function is not a quadratic function of the states and inputs. Additionally, it is crucial to impose constraints on the inputs to ensure they adhere to the physical limitations of engine operation and driving comfort. Furthermore, Eq. (\ref{eq:SoC}) imposes a complex nonlinear constraint of states and inputs on the system, i.e., $(P_{batt}(k)\leqslant \frac{V^{2}_{batt}}{4 R_{batt} })$. To the best of the authors' knowledge, none of the current ADP algorithms can address this type of nonlinear constraint.

\begin{figure}[!t]

 \center

  \includegraphics[draft=false, width=3.8in]{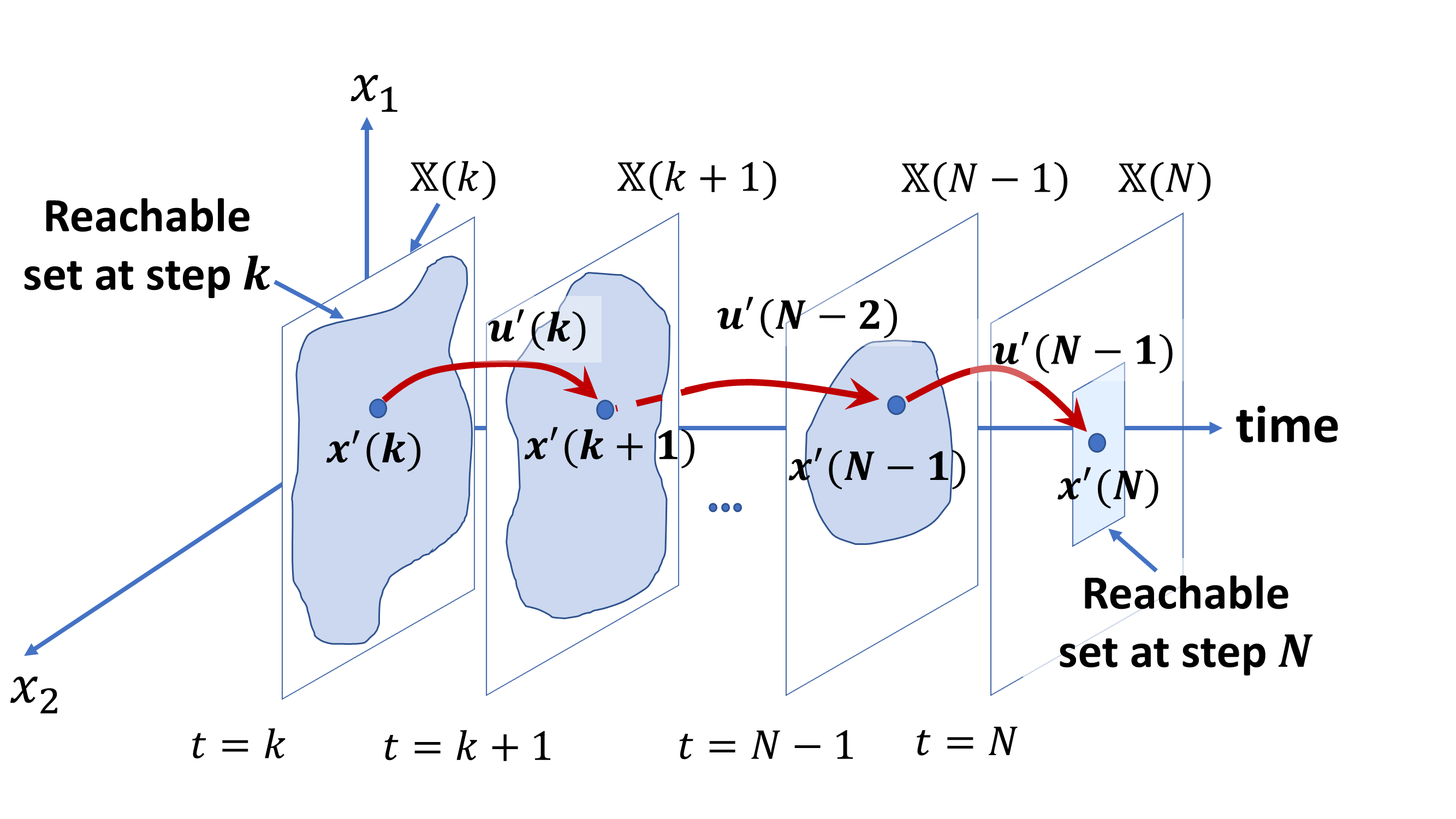}\\
  \caption{Illustration of Reachable Sets.}\label{fig:ReachableSet}
\end{figure}

To address these issues, in this study, we propose to modify the original ADP method to offer a customized ADP method for the optimization problem considered. First of all, this customized ADP method is applicable to both control-affine and non-affine systems. Furthermore, it is capable of handling non-quadratic nonlinear cost functions associated with both states and inputs. Finally, it adopts a way to handle complex constraints using the concept of reachable sets. 

Particulary, in this method, the approximated optimal cost-to-go $\bar{V}^{*}_{k}(\textbf{X}(k))$ at each step is determined through the utilization of deep neural networks (DNNs):
\begin{equation} \label{eq:V_bar_star_X}
\bar{V}^{*}_{k}(\textbf{X}(k)) = \phi_{k}(\textbf{X}(k)). 
\end{equation}
In this DNN, the input consists of the current state $\textbf{X}(k)$, and the output is the approximation of the optimal cost-to-go $\bar{V}^{*}{k}(\textbf{X}(k))$. The architecture of the DNN, denoted as $\phi{k}(.)$, is illustrated in Figure \ref{fig:NN}. To detail this method, first, the concept of reachable sets needs to be addressed.

\subsection{Reachable Set}

To address the complex constraints associated with the optimization problem, the concept of reachable sets is adopted to the customized ADP framework. The reachable set, at each step $k$, is defined as the set of points for which there is at least one sequence of actions under which the system transforms from the current state to a state in the desired region at the end of the time horizon. Fig. \ref{fig:ReachableSet} illustrates the concept of reachable sets for a two-dimensional dynamical system. For instance, in this figure, $\textbf{x}^{'}(k)$ is a point in the reachable set at step $k$ since there is at least a sequence of actions $\{ \textbf{u}^{'}(k), ...,  \textbf{u}^{'}(N-1) \}$ which can transform $\textbf{x}^{'}(k)$ to a point $\textbf{x}^{'}(N)$ in the desired region at step $N$.
Note that this sequence of actions may not necessarily be the optimal sequence of actions from $\textbf{x}^{'}(k)$ to the end of the horizon. 
In this figure, each rectangle shows the horizon of the state space at time index $k$, and each blue region shows the reachable set at time index $k$. In order to find the reachable set at each step, we start backward in time. 

Specifically, for the system defined in \ref{State_Space}, the reachable sets are obtained as follows: for $k=N$, the reachable set is defined as the set containing all the desired terminal points $\textbf{X}_{des}(N)$:
\begin{align}
	\nonumber
	\mathcalligra{R}{(N)} 
	&= 
	\{
	\textbf{X}_{des}(N)
	\in 
	\mathbb{R}^{3} 
	\mid
	\\  
	\textbf{X}_{des, min}(N)
	&\leqslant
	\textbf{X}_{des}(N)
	\leqslant
	\textbf{X}_{des, max}(N) 
	\}
	\label{eq:ReachableSet_def}
\end{align}
where $\textbf{X}_{des, min}(N)$ and $\textbf{X}_{des, max}(N)$ denote the minimum and the maximum range for $\textbf{X}_{des}(N)$, respectively. 
Given the constraints on control inputs and the constraints set by the system dynamics, i.e., $\textbf{X}(k) \in\mathbb{X}(k)$ and $\textbf{U}^{0}(k) \in	\mathbb{U}(k)$,  the reachable sets from step $N-1$ to step 1 are found following the method in \cite{elbert2012implementation} as:

\begin{align} \label{eq:ReachableSet}
	&\mathcalligra{R}~(k) 
	= 
	\{
		\textbf{X}(k)
		%\in
		%\boldsymbol{\mathscr{X}}(k)
		\mid
		\exists ~ 
		\textbf{u}^{0}(k)
		\nonumber
	\\
	& \text{s.t.} 
	\left\{
	\begin{aligned}
		\textbf{X}(k+1) 
		&= 
		\textbf{X}({k}) + \delta t \mathscr{F}(\textbf{X}(k), \textbf{u}^{0}(k))
		\in 
		\mathcalligra{R}~(k+1)
	\\
	\textbf{X}(k)
	&\in
	\mathbb{X}(k)	
	\\
	\textbf{u}^{0}(k)
	&\in
	\mathbb{U}(k)	
	\\		
	P_{batt}(k)
	&\leqslant
	\frac
	{
	V^{2}_{batt}	
	}
	{
	4 R_{batt} 
	}		
%	\nonumber
\end{aligned}
	\right.
	\nonumber
\\
& \qquad\qquad\quad\}.
\end{align}
%
%$$$$$$$$$$$$$$$$$$$$$$$$$$$$

%
%
%
%
%
The constraints specified in Eq. (\ref{eq:ReachableSet}) are imposed to ensure that the states and inputs of the system remain within the feasible region. For instance, in the case of HEVs, the $SoC$ must be within the range of 0 to 1, and the engine speed ${\omega}_{e}$ should not exceed its maximum value (approximately 450 rad/s) or be negative. Furthermore, the last constraint in Eq. (\ref{eq:ReachableSet}) is introduced to guarantee that $SoC$ remains a real number.

The time-variant state grid $\mathbb{X}(k)$ and input grid $\mathbb{U}(k)$ can be defined as follows:
\begin{align}\label{eq:Xgrid}
	\mathbb{X}(k) 
	= 
	\{
	\textbf{X}(k)
	\mid
	\textbf{X}_{min}(k)
	\leqslant
	\textbf{X}(k)
	\leqslant
	\textbf{X}_{max}	(k)
	\}
	\\
	\label{eq:Ugrid}
	\mathbb{U}(k) 
	= 
	\{
	\textbf{U}(k)
	\mid
	\textbf{U}_{min}(k)
	\leqslant
	\textbf{U}(k)
	\leqslant
	\textbf{U}_{max}	(k)
	\}
\end{align}
where ($\textbf{X}_{min}(k)$, $\textbf{X}_{max}(k)$), and ($\textbf{U}_{min}(k)$, $\textbf{U}_{max}(k)$) denote the minimum and the maximum limits for $\textbf{X}(k)$ and $\textbf{U}(k)$, respectively. 
%
%
%
%
%
%
%The last constraint in (Eq. 29)is to make sure that the $SoC$ of the battery is a real number throughout the drive cycle. 
Besides, the control input by which $V_{k}^{*}(\textbf{X}(k))$ is attained is the optimal control input, $\textbf{U}^{*}(\textbf{X}(k))$. 
\subsection{ADP training within the reachable set}
Once the reachable set is identified, the training of the customized ADP method is based on the reachable set. Unlike the standard ADP in which the region of training is not restricted, in the customized ADP used in this paper, the training is only conducted within the reachable set.  In this way, it can be ensured that the optimal control solution identified can always meet constraints of the optimization problem. Specifically, only the points inside the reachable set $\mathcalligra{R}~(k)$ are used to train the network $\phi_{k}(.)$. Once the network is trained, it can be used to approximate the cost-to-go within the desired set. The training procedure is done backward in time. At first, the network learns the optimal cost-to-go at the final step using Eq. (\ref{eq:Vstar_XN}) for the points inside $\mathcalligra{R}{(N)}$, i.e., $\phi_{N}(.)$ approximates $V_{N}^{*}(.)$. Then, at each step $k \neq N$, $h$ number of random training samples inside $\mathcalligra{R}{(k)}$ are chosen. Subsequently, the optimal control input ${\textbf{U}^{*}(k)}$ is required to determine the approximated minimum cost-to-go $\bar{V}^{*}_{k}(.)$. In the standard ADP methods \cite{heydari2012finite, li2021novel, kiumarsi2019optimal, abu2005nearly}, the cost-to-go is defined in a quadratic form and the system is control affine (Eq. (\ref{eq:X_affine})); thus, this step is solved analytically through Eq. (\ref{eq:u_star_affine}). However, for the particular optimization problem we considered in this paper, the cost function is not quadratic and the system is not control affine (Eq. 32). To still be able to find the optimal control inputs under this setting, we replace the analytical solution in Eq. (34) by a numerical process shown as Steps 3 and 4 in the Table of Algorithm \ref{alg:1}. This is to compare the costs associated with different values of the control inputs within the allowable range and choose that with the minimum cost. Note that this numerical step won't cause much time consumption since the dimension of the input space is low. Once this step is done for all the sample points, the sample state points and their corresponding $\bar{V}^{*}_{k}(.)$ will be used as the features and the targets to train the neural network $\phi_{k}(.)$, i.e., steps 5-10 in the Table of Algorithm\ref{alg:1}.
\begin{align} \label{eq:J_star_Xk}
	&\bar{V}^{*}(\textbf{X}(k)) 
	= 
	\phi_{k}(\textbf{X}(k))
	\nonumber
	\\ 
	&= 
		\begin{cases}
			\psi(\textbf{X}(N), \qquad for \quad k=N\quad \& \quad \textbf{X}(N) \in \mathcalligra{R}{(N)} 
			\\
			\\
			\begin{aligned}
			&
			\minA\limits_
			{\textbf{U}(k) \in \mathbb{U}(k)	} 
			{
			\big(\delta t \left (\dot{m}_{fuel}(k)\right) 
			+
			\phi_{k+1}(\textbf{X}(k+1))\big)
			},  
			\\
			&\qquad\qquad\qquad for \quad k\neq N \quad \& \quad \textbf{X}(k) \in \mathcalligra{R}(k).
			\end{aligned}
		\end{cases}
\end{align}

After completing the training process, the trained neural network is utilized to obtain the sub-optimal constrained control input sequence given the initial condition $\textbf{X}(0)$. The implementation algorithm is further described in Algorithm \ref{alg:2}.
%
%
%
%
%\begin{remark}\label{remark:1}
%Hyperbolic tangent sigmoid transfer function is selected as the transfer function between the layers. Besides, the gradient-based method Bayesian regularization backpropagation is used to train the NN. The training algorithm is detailed in algorithm \ref{alg:1}.
%\end{remark}
\begin{algorithm}
 \caption{Training Neural Network}
 \label{alg:1}
\SetAlgoLined
\nl Select a small positive number $\alpha$, and a big enough integer $IterMax$\;
 \nl 
	\For{
	{k} 
	= 
	{N}
	:  
	{-1}
	:  
	{0}
	}
	{
 		\nl 
		Choose $h$ different random training samples  
		$\textbf{X}^{l}(k)$ 
		in the reachable set 
		$\mathcalligra{R}~(k)$ 
		where 
		$l \in \{1,2,..., h\}$\;
	 	\nl
		For each training sample 
		$\textbf{X}^{l}(k)$
		find $\bar{V}_{k}^{*}(\textbf{X}^{l}(k))$ using Eq. (\ref{eq:J_star_Xk})
		\;
		\nl
		Initialize 
		$\phi^{0}_{k}(.)$
		with random parameters\;
 		\nl 
		\For
		{
		\textnormal{i} 
		= 
		\textnormal{1}
		: 
		$IterMax$
		}
			{
				\nl 
				Update the neural network $\phi^{i}_{k}(.)$ and find the parameters to approximate
				$\bar{V}_{k}^{*}(\textbf{X}^{l}(k))$ 
				using backpropagation on the entire training samples\;
					\nl 
					\If
					{
						${\Vert \phi^{i}_{k}(.) - \phi^{i-1}_{k}(.) \Vert \leq \alpha}$
					}
						{
						   \nl 
							\textbf{Break}\;
						}
		    }
			\nl 
			$\phi_{k}(.)$ 
			$\leftarrow$
			$\phi^{i}_{k}(.)$\;
    }
\end{algorithm}
\begin{algorithm}
 \caption{Implementation}
 \label{alg:2}
\SetAlgoLined
	\nl 
	\For
	{
	{k} 
	= 
	{0}
	: 
	$N-1$
	}
		{
		 %\begin{align*}
		\nl
		$\textbf{U}^{*}(\textbf{X}(k))=$
			\begin{equation*}
			\argminB\limits_
			{\textbf{U}(k) \in \mathbb{U}(k)	} 
			%{
			\big(\delta t  (\dot{m}_{fuel}(k) ) 
			+
			 \phi_{k+1}(\textbf{X}(k+1))\big);
			%}
			\end{equation*}
			% Note that {aligned} environment cannot be used inside an algorithm
		%$
		\nl
 		$
		 \textbf{X}(k+1) = \textbf{X}({k}) + \delta t \mathscr{F}(\textbf{X}(k), \textbf{U}(k)) 
		$
	 }	
\end{algorithm}
\section{Simulation Results} \label{Results} 
\begin{figure}[t]  % t,h, and b mean "top," "in-line," and "bottom" respectively
  \centering
  \includegraphics[draft=false, width=3.8in]{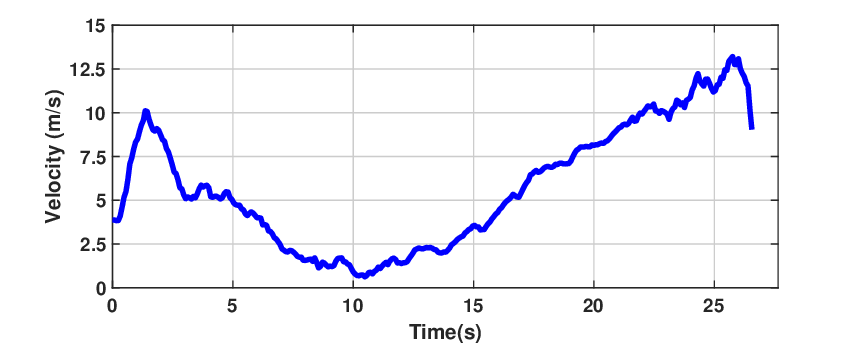}\\
  \caption{Target Vehicle Velocity Profile.}\label{fig:Velocity_Profile}
%\end{figure}

%\begin{figure}[!htb] 
%\centering

\bigskip

\includegraphics[draft=false, width=3.8in]{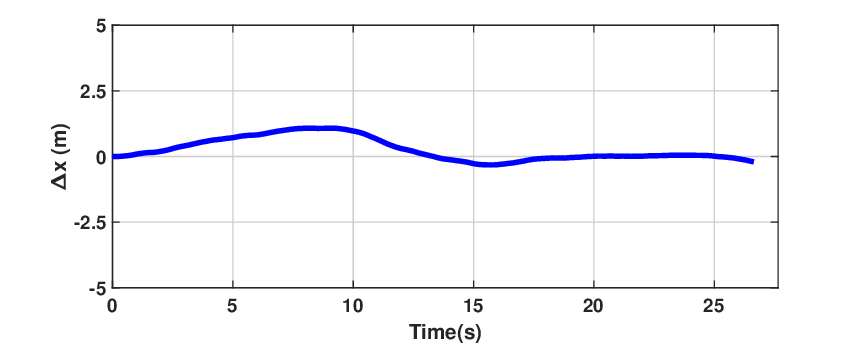}\\
  \caption{Displacement Deviation History for the Vehicle.}\label{fig:DX}

\bigskip

\includegraphics[draft=false, width=3.8in]{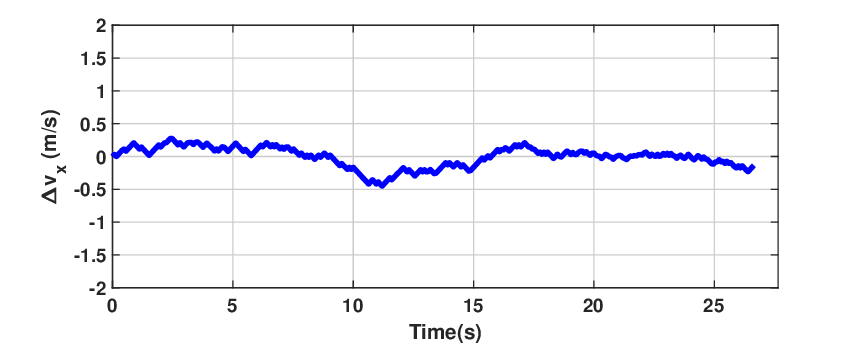}\\
  \caption{Velocity Deviation History for the Vehicle.}\label{fig:DV}

%\bigskip
%  \includegraphics[draft=false, width=3.8in]{pictures/SoC}\\
%  \caption{Battery State of Charge History of the Vehicle.}\label{fig:SoC}

\end{figure}

In this section, the performance of the proposed controller is evaluated using a real-world data set \cite{khajeh2021novel}. The data set contains recorded information on the external kinematics of vehicles, such as velocity, acceleration, and headway to leading and rear vehicles. The data was collected on a 150-meter-long section of the I-35 Corridor in Austin, USA. To assess the effectiveness of the controller, a random vehicle from the data set is selected. Fig. \ref{fig:Velocity_Profile} displays the velocity profile of the target vehicle.

%
%\begin{figure}[!h]  % t,h, and b mean "top," "in-line," and "bottom" respectively
%  \centering
%  \includegraphics[draft=false, width=3.8in]{pictures/Relative_Dist}\\
%  \caption{Relative Distance of the neighbor vehicles.}\label{fig:Relative_Dist}
%\end{figure}
To ensure passenger safety and driving comfort, the corresponding constraint Eqs. (\ref{eq:limitations_x}) to (\ref{eq:limitations_T}) are enforced on $\Delta{x}$, $\Delta{v}$, and $\Delta{T}_{d}$. These constraints restrict the maximum range of flexibility in longitudinal displacement $\Delta{x}$ and velocity $\Delta{v}$ based on the relative distances and velocities with the front and rear vehicles at each step. The flexibility in driveline torque $\Delta{T}_{d}$ is constrained by the maximum torque of the engine and motor at the current speed. For simplicity, the limitations of these flexibilities remain constant \cite{zhang2021adaptive, kargar2022optimal}.
\begin{align}
    \label{eq:limitations_x}
	\Delta{x}_{min}
	&\leqslant
	\Delta{x}(t)
	\leqslant
	\Delta{x}_{max}
	\\  
	\label{eq:limitations_v}
	\Delta{v}_{min}
	&\leqslant
	\Delta{v}(t)
	\leqslant
	\Delta{v}_{max}
	\\  
	\label{eq:limitations_T}
	\Delta{T}_{d,min}
	&\leqslant
	\Delta{T_d}(t)
	\leqslant
	\Delta{T}_{d,max}	
\end{align}
Then, the penalizing function $\psi(\textbf{X}(N)$ should be designed which is used to ensure that the vehicle will reach a point in the desired terminal set. As mentioned, the controller is trained using the points inside the corresponding reachable set. Thus, a narrow desired terminal set is not favorable from the training perspective as the DNN cannot be well trained. Thus, the desired terminal set must be broad enough, and at the same time, must favor the points close to the ideal terminal conditions, i.e., $SoC_{ideal}(N) = SoC(0), \Delta{x}_{ideal}(N)=0, \Delta{v}_{ideal}(N) = 0$. We define $\psi(\textbf{X}(N)$ as follows:
\begin{align} \label{eq:psi}
	\psi(\textbf{X}(N))	 
	&
	\triangleq 
	max (c_{1}, c_{2}, c_{3}) 
\end{align} 
where
\begin{align} \label{eq:ci}
	&c_{i}= 
		\begin{cases}
			\begin{aligned}
			%&\gamma  &x_{i} < x_{i, min}(N)
			%\\
			&\gamma 
			\frac
			{x_{i, low} - x_{i}}
			{x_{i, low} - x_{i, min}(N)}   &x_{i, min}(N)\le x_{i} < x_{i, low}
			\\
			&0   &x_{i, low}\le x_{i} \le x_{i, high}
			\\
			&\gamma 
			\frac
			{x_{i} - x_{i, high}}
			{x_{i, max}(N) - x_{i, high}} &  x_{i, high}< x_{i} \le x_{i, max}(N)
			%\\
			%&\gamma  &  x_{i, max}(N)<x_{i}
			\end{aligned}
		\end{cases}
\nonumber
	\\
	&\qquad i=\{1,2,3\}
\end{align}
where $x_{1}$, $x_{2}$, and $x_{3}$ represent $\Delta{x}$, $\Delta{v}$, and $SoC$, respectively. Also, $x_{i, min}(N)$ and $x_{i, max}(N)$ represent the lower and upper bounds of the desired terminal set, and ${x_{i, low}}$, and ${x_{i, high}}$ represent the limits of the region in which the cost is zero. For the points outside that region, the cost linearly increases to the constant value $\gamma$.
\begin{figure}[!htb] 
\centering

%\includegraphics[draft=false, width=3.8in]{pictures/DX}\\
%  \caption{Displacement Deviation History for the Vehicle.}\label{fig:DX}

%\bigskip

%\includegraphics[draft=false, width=3.8in]{pictures/DV}\\
%  \caption{Velocity Deviation History for the Vehicle.}\label{fig:DV}

%\bigskip

  \includegraphics[draft=false, width=3.8in]{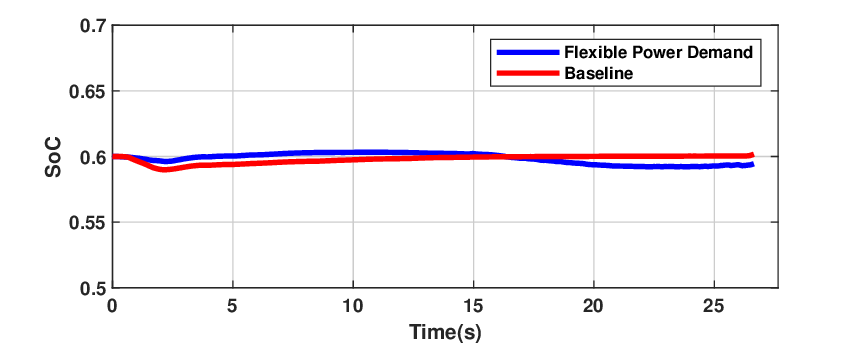}\\
  \caption{Battery State of Charge History of the Vehicle.}\label{fig:SoC}

\bigskip

\includegraphics[draft=false, width=3.8in]{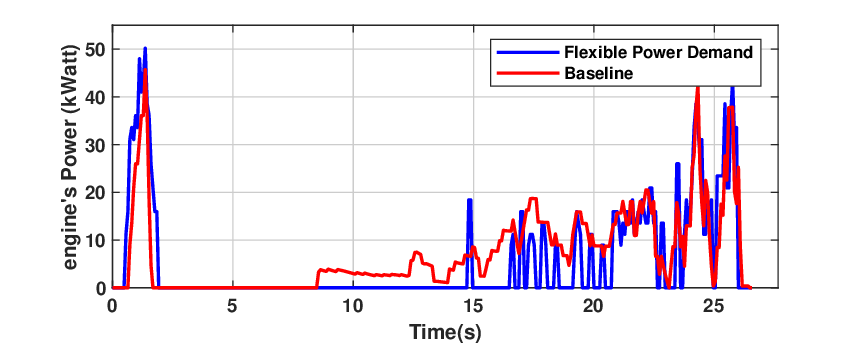}\\
  \caption{Engine Power History of the Vehicle.}\label{fig:Pe}

\bigskip

  \includegraphics[draft=false, width=3.8in]{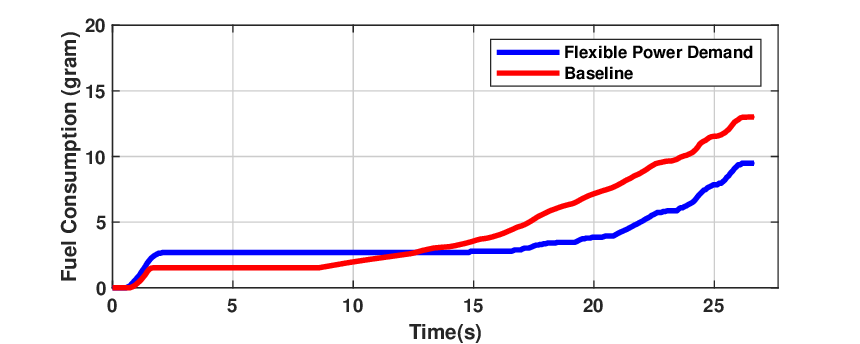}\\
  \caption{Fuel Consumption History of the Vehicle.}\label{fig:Fuel}

\end{figure}

The control performance of the system utilizing the proposed control method is evaluated for an initial condition $\textbf{X}(0)= [0(m), 0(m/s), 60\%]^T$. The performance results are depicted in Figures (\ref{fig:DX}) - (\ref{fig:Fuel}). To provide a basis for comparison, the performance of the system under the baseline approach is also illustrated in Figs. (\ref{fig:SoC}) - (\ref{fig:Fuel}). In the baseline optimization, the flexibility in power demand is not considered (Fixed Demand), and the controller's task is to supply the precise amount of power required by the upper-level dynamics. Consequently, in the baseline optimization, the state space is reduced to [$SoC$], and the input space consists of [$\omega_{eng}$].

Fig. \ref{fig:Fuel} shows fuel consumption histories using both methods. In practice, the terminal $SoC$ may slightly differ from the terminal desired $SoC$. Following \cite{onori2016hybrid}, a fuel consumption compensation method is used in this paper (Eq. (\ref{eq:fuel_compensation})) to consider the final $SoC$ deviation from its desired value.
\begin{equation} \label{eq:fuel_compensation}
Fuel_{comp} = Fuel_{act}-\kappa \Delta{SoC}
\end{equation}
where $Fuel_{comp} $, $Fuel_{act}$, and $\Delta{SoC}$ represent the compensated fuel consumption corresponding to a zero $SoC$ deviation, the actual fuel consumption, and the final $SoC$ deviation, respectively. Parameter $\kappa$, which converts $\Delta{SoC}$ into a corresponding amount of fuel, is a curve-fitting coefficient \cite{onori2016hybrid}. Comparisons between the compensated fuel consumption for the strategies shown in Table \ref{table:1} displays that an additional 14.1\% fuel economy improvement was achieved by the proposed algorithm over the baseline (Fixed Demand) strategy.

\begin{table}[!h]
\centering
\caption{Summary of Results for the Drive Cycle}
\label{table:1}
\begin{tabularx}{3.45in}{C | C | C | C | C}
\toprule
       Method & Terminal $\Delta {x}$ (m) & Terminal $\Delta {v}$ (m/s) & Terminal $SoC$ (\%)  & $Fuel_{comp}$ (g)\\ 
\midrule
        Flexible Demand    & -0.20       & -0.14      & 59.45       &  10.81  \\
\midrule
        Fixed Demand    & $-$       & $-$       & 60.17             & 12.59   \\
%%\midrule
%%        $X_{max,2}$    & 10 $\frac{m}{s}$       & $X_{max,3}$       & 100 $percent$\\
%%\midrule
%%        $X_{max,4}$    & 180 $(N.m)$       & $X_{max,5}$       & 10 $\frac{rad}{s}$\\
\bottomrule
\end{tabularx}
\end{table}

%

%\begin{figure}[!t] 
%\centering

%\includegraphics[draft=false, width=3.8in]{pictures/Pe}\\
%  \caption{Engine Power History of the Vehicle.}\label{fig:Pe}

%\bigskip

%  \includegraphics[draft=false, width=3.8in]{pictures/Fuel}\\
%  \caption{Fuel Consumption History of the Vehicle.}\label{fig:Fuel}

%\end{figure}

%
%
%
\section{Conclusion} \label{Conclusion} 
Traditionally, vehicle coordination optimization and powertrain energy management have been studied separately in the context of hybrid electric vehicles (HEVs). However, this paper introduces a novel approach that combines and optimizes these two levels simultaneously by leveraging a unique feature which exists in autonomous HEVs: flexible power demand. The concept of flexible power demand acknowledges that the powertrain does not necessarily have to meet the power required by the external dynamics of an autonomous HEV at every step. By exploiting this flexibility, the proposed powertrain energy management method aims to enhance fuel economy.

The optimization problem is formulated within the framework of a customized approximate dynamic programming method, utilizing the notion of reachable sets. To evaluate the effectiveness of the proposed method, a case study is conducted using real-world data. The results demonstrate a significant 14.1 \% improvement in fuel consumption compared to a conventional optimization method that employs fixed power demand. This highlights the potential of the proposed approach in achieving enhanced fuel efficiency in HEVs.
%

%\begin{table}[!h]
%\centering
%\caption{Summary of Results for the Drive Cycle}
%\label{table:1}
%\begin{tabularx}{3.45in}{C | C | C | C | C}
%\toprule
%       Method & Terminal $\Delta {x}$ (m) & Terminal $\Delta {v}$ (m/s) & Terminal $SoC$ (\%)  & $Fuel_{comp}$ (g)\\ 
%\midrule
%        Flexible Demand    & -0.20       & -0.14      & 59.45       &  10.81  \\
%\midrule
%        Fixed Demand    & $-$       & $-$       & 60.17             & 12.59   \\
%%\midrule
%%        $X_{max,2}$    & 10 $\frac{m}{s}$       & $X_{max,3}$       & 100 $percent$\\
%%\midrule
%%        $X_{max,4}$    & 180 $(N.m)$       & $X_{max,5}$       & 10 $\frac{rad}{s}$\\
%\bottomrule
%\end{tabularx}
%\end{table}
% if have a single appendix:
%\appendix[Proof of the Zonklar Equations]
% or
%\appendix  % for no appendix heading
% do not use \section anymore after \appendix, only \section*
% is possibly needed

% use appendices with more than one appendix
% then use \section to start each appendix
% you must declare a \section before using any
% \subsection or using \label (\appendices by itself
% starts a section numbered zero.)
%

\appendices
% you can choose not to have a title for an appendix
% if you want by leaving the argument blank
\section{}  \label{Appendix}
%Below is the environment specifications \& vehicle system parameters:

\begin{table}[!ht]
\renewcommand\thetable{A.\rom{1}}
\centering
\caption{Environment Specifications \& Vehicle System Parameters}
\label{table:2}
\begin{tabularx}{6.9in}{C C | C C | C C | C C}
\toprule
        Parameter & Value & Parameter & Value &Parameter & Value & Parameter & Value\\ 
\midrule
        m   & 1350 kg      & r     & 0.28 m 	&  $r_r$   & 0.078 m      & $r_s$       & 0.030 m\\
\midrule
        ${\mu}_R$   & 0.007      & $\rho$      & 1.225 kgs/m$^3$ 	&$C_d$   & 0.3      & $A_f$      & 2.2 m$^2$\\
\midrule
        $k_c$  &  3.9      &$\mu_{g}$      & 0.9 	&$\mu_{m}$   & 0.9       &$V_b$       & 202 V\\
\midrule
         $Q_b$    &  23400 A.s       &$R_b$      & 0.45 $\Omega$ 	&$\Delta{x}_{min}$    &  -3.5 m       &$\Delta{x}_{max}$      & 3.5 m\\
\midrule
         $\Delta{v}_{min}$    &  -2.5 m/s       &$\Delta{v}_{max}$      & 2.5 m/s 		&$SoC_{min}$    &  50\%       &$SoC_{max}$      & 70\%\\
\midrule
         $\Delta{T}_{d,min}$    &  -150 N.m     &$\Delta{T}_{d,max}$      & 150 N.m 	&$\omega_{e, min}$    &  0 rad/s       &$\omega_{e, min}$      & 450 rad/s\\
\midrule
         $\Delta{x}_{min}(N)$    &  -2.0 m       &$\Delta{x}_{max}(N)$      & 2.0 m 		&$\Delta{x}_{low}$    &  -0.5 m       &$\Delta{x}_{high}$      & 0.5 m\\
\midrule
         $\Delta{v}_{min}(N)$    &  -1.5 m/s       &$\Delta{v}_{max}(N)$      & 1.5 m/s 	&$\Delta{v}_{low}$    &  -0.5 m/s       &$\Delta{v}_{high}$      & 0.5 m/s \\
\midrule
         $SoC_{min}(N)$    &  53\%      &$SoC_{max}(N)$      & 67\% 		&$SoC_{low}$    &  59\%       &$SoC_{high}$      & 63\%\\
\bottomrule
\end{tabularx}
\end{table}

%
%
%
%
%
%
%
%
%
%

% you can choose not to have a title for an appendix
% if you want by leaving the argument blank
%%\section{}
%%Appendix two text goes here.

% use section* for acknowledgment
\begin{comment}
\section*{Acknowledgment}
This research was partially supported by the National Science Foundation under Grant No. 1826410.
\end{comment}
%
% Can use something like this to put references on a page
% by themselves when using endfloat and the captionsoff option.
\ifCLASSOPTIONcaptionsoff
  \newpage
\fi

% trigger a \newpage just before the given reference
% number - used to balance the columns on the last page
% adjust value as needed - may need to be readjusted if
% the document is modified later
%\IEEEtriggeratref{8}
% The "triggered" command can be changed if desired:
%\IEEEtriggercmd{\enlargethispage{-5in}}

% references section

% can use a bibliography generated by BibTeX as a .bbl file
% BibTeX documentation can be easily obtained at:
% http://mirror.ctan.org/biblio/bibtex/contrib/doc/
% The IEEEtran BibTeX style support page is at:
% http://www.michaelshell.org/tex/ieeetran/bibtex/
%\bibliographystyle{IEEEtran}
% argument is your BibTeX string definitions and bibliography database(s)
%\bibliography{IEEEabrv,../bib/paper}
%
% <OR> manually copy in the resultant .bbl file
% set second argument of \begin to the number of references
% (used to reserve space for the reference number labels box)
\begin{comment}

\end{comment}
\bibliographystyle{IEEEtran}
\bibliography{AHEV_PowerFlex}
% biography section
% 
% If you have an EPS/PDF photo (graphicx package needed) extra braces are
% needed around the contents of the optional argument to biography to prevent
% the LaTeX parser from getting confused when it sees the complicated
% \includegraphics command within an optional argument. (You could create
% your own custom macro containing the \includegraphics command to make things
% simpler here.)
%\begin{IEEEbiography}[{\includegraphics[width=1in,height=1.25in,clip,keepaspectratio]{mshell}}]{Michael Shell}
% or if you just want to reserve a space for a photo:

\begin{comment}
% if you will not have a photo at all:
\begin{IEEEbiographynophoto}{John Doe}
Biography text here.
\end{IEEEbiographynophoto}

% insert where needed to balance the two columns on the last page with
% biographies
%\newpage
\end{comment}

\begin{comment}
\begin{IEEEbiographynophoto}{Jane Doe}
Biography text here.
\end{IEEEbiographynophoto}
\end{comment}
% You can push biographies down or up by placing
% a \vfill before or after them. The appropriate
% use of \vfill depends on what kind of text is
% on the last page and whether or not the columns
% are being equalized.

%\vfill

% Can be used to pull up biographies so that the bottom of the last one
% is flush with the other column.
%\enlargethispage{-5in}

% that's all folks
\end{document}